\documentclass{article}  
\usepackage{breckenridge}
\usepackage{epsfig,graphicx}
\frompage{000} \topage{000}           
\def\rightmark{Strangeness Saturation}
\def\leftmark{B. K\"ampfer et al.}
 
\title{Strangeness Saturation:\\
Dependence on System-Size, Centrality and Energy$^a$} 
\authors{
{B. K\"ampfer$^1$,
J. Cleymans$^2$, 
P. Steinberg$^{2,b}$
and
S. Wheaton$^2$ 
}\\[2.812mm]
{\normalsize
\hspace*{-8pt}$^1$ Forschungszentrum Rossendorf, PF 510119,
01314 Dresden, Germany\\[0.2ex] 
\hspace*{-8pt}$^2$ Department of Physics, University of Cape Town, \\
Rondebosch 7701, Cape Town, South Africa
}}
 
\abstract{The dependence of the strangeness saturation factor on the
system size, centrality and energy is studied in relativistic
heavy-ion collisions.}

 
\begin{document}
 
\maketitle
\setcounter{page}{1}

\section{Introduction}\label{intro}

Statistical-thermal models (see \cite{review} for a review) 
aim at reproducing hadron multiplicities
in relativistic heavy-ion collisions in certain phase space regions by a small number
of parameters. As these parameters evolve smoothly with external conditions, thermal
models have predictive power. Moreover, the thermal parameters, taken literally,
allow for a quantification of the conditions reached in the transient stage 
of compressed and heated strongly interacting matter. While there is exhaustive literature
on the successful use of the thermal models, the foundation of the applicability
is far from settled.
One should also keep in mind that thermal models address only a subset of the wealth of
observables.

The present contribution summarizes our recent work \cite{Jean,Jean_submit,bk} investigating the 
dependence of strangeness saturation on system size, centrality and energy, both in full
phase space ($4\pi$) and at mid-rapidity ($y \approx 0$).

\section{Thermal Model}

Our model is based on the grand canonical ensemble expression for the abundance
of hadron species $i$ (Fermions [+1] or Bosons [-1]) with degeneracy $g_i$
\begin{equation}
N_i^{\rm prim} = V g_i
\int \frac{d^3 p}{(2\pi)^3} \, dm_i \,
\left[ \gamma_s^{ - \left| S_i \right| } 
\exp \left\{ - \frac{E_i - \vec \mu_i \vec Q_i}{T} \right\} \pm 1 \right]^{-1}
\mbox{BW} (m_i, \Gamma_i),
\end{equation}
where $T$ is the temperature, $\vec \mu$ a set of chemical potentials corresponding
to the conserved charges $\vec Q$; $E_i = \sqrt{ \vec p \, ^2 + m_i^2 }$ with vacuum
mass $m_i$.

For small particle numbers one has to turn to a canonical ensemble, e.g.
by the projection method \cite{projection}. It is important to include resonances
(we include meson (baryon) states composed of the three lightest quarks and anti-quarks 
up to 2.3 (2.6) GeV); 
their widths are taken into account via
the Breit-Wigner parameterization $BW(m_i, \Gamma_i)$, which collapses
to a $\delta$-function for the respective ground states. Due to feeding and resonance
decays, the final hadron multiplicities become
$N_i = N_i^{\rm prim} + 
\sum_j \left\{ N_j^{\rm prim} Br (j \to i) - N_i^{\rm prim} Br (i \to j)] \right\}$,
where $Br$ denotes the corresponding branching ratio.

Our focus here is the phenomenological strangeness saturation factor $\gamma_s$ in Eq.~(1).
It is thought \cite{Rafelski} to parameterize possible deviations from chemical equilibrium 
in the strange sector. Whether one common factor $\gamma_s$ is sufficient,
with powers $\vert S_i \vert$ determined by the
total strangeness content of hadron $i$, needs still to be tested by analyzing data.
Also in the non-strange sector such an off-equilibrium parameter should be
introduced \cite{Jan}. Due to the small data samples at our disposal, such an additional
parameter cannot be unambiguously fixed; therefore, we are forced here to assume full
equilibration in the non-strange sector.

The parameters $T$ and $\vec \mu$ appear exponentially in Eq.~(1), while the sensitivity
on variations of $\gamma_s$ is weaker. For this reason, we study the mentioned parameter 
dependencies based on analyses of data sets comprised of the same hadron species. Otherwise 
variations of $\gamma_s$ can be absorbed in small changes of $T$ and $\vec \mu$. Unfortunately, 
this shrinks the wealth of available data to quite poor subsets. 

Within the Cooper-Frye formalism, dynamical effects factor out of fully-integrated hadron 
yields \cite{Cooper_Frye}, provided that freeze-out occurs on a hyper-surface with constant $T$ 
and $\vec \mu$. In this respect, Eq.~(1) covers also dynamical
situations. 
Instead of considering hadron ratios, 
we consider multiplicities $N_i^{\rm prim}$ with a fiducial normalization volume $V$ in Eq.~(1).

\section{Data Analysis}
\subsection{CERN-SPS energies}
\subsubsection{Centrality and System Size Dependence 
at $E_{\rm beam} = 158$ AGeV}

$4\pi$ multiplicities of $\pi^\pm$, $K^\pm$, $\phi$ and $N_{\rm part}$ are at our disposal for
(i) central collisions of C + C and Si + Si, and
(ii) centrality binned collisions Pb + Pb. More baryon information is desirable, but only
$\bar p$'s are available additionaly for Pb + Pb. The left panel of 
Fig.~1 summarizes our findings \cite{Jean_submit}. We observe,\\
$\bullet$ Pb + Pb: $\gamma_s$ increases with centrality, but stays below unity, irrespective
of whether $\bar p$'s are included or not; inclusion of $\bar p$'s reduces $\gamma_s$ somewhat; 
minimizing $\chi^2$ or the quadratic deviation of data and the model multiplicities, $q^2$, 
yields slightly different results.\\
$\bullet$ C + C, Si + Si: the extracted values of $\gamma_s$ off-set the down-extrapolation
for Pb + Pb. This again emphasizes the findings of NA49 \cite{C-Si}: system-size and centrality
dependencies differ.\\
$\bullet$ Most striking is that $\gamma_s$ goes as the number of participants which underwent
multiple collisions, as extracted from a Glauber model calculation (cf.\ Fig.~3
in \cite{Jean_submit}). 

There is some controversy on the use of $4\pi$ or mid-rapidity data. Our analysis 
\cite{Jean_submit} of $\pi^\pm$, $K^\pm$ and $p^\pm$ centrality-binned mid-rapidity data
from Pb + Pb at 158 AGeV reveals that $\gamma_s$ does indeed differ for $4\pi$ and 
$y \approx 0$ data (see Fig.~1, middle panel). This is due to the fact that the rapidity 
distributions of various hadrons differ. Combining the results, even the very restricted data 
samples point to $\gamma_s = 1 \pm 0.25$ for most central collisions. This suggest that 
$\gamma_s = 1$ with slightly renormalized values of $T$ and $\vec \mu$ 
is a good choice for the given beam energy and high centrality. 

\subsubsection{Energy Dependence}

Unfortunately, the previous analysis cannot be extended to extract energy dependencies.
Rather, the $4 \pi$ multiplicities of $\pi^\pm$, $K^\pm$, $\Lambda$ and $\bar \Lambda$
at beam energies of 40, 80 and 158 AGeV are at our disposal as comparable data samples.
The results of our analysis \cite{Jean_submit} are shown in Fig.~2: $\gamma_s$
displays a dip at 80 AGeV,
while $\gamma_s(40 {\rm AGeV}) \sim \gamma_s(158 {\rm AGeV})$. 
(The significance of the drop at 80 AGeV still needs to be clarified.)
$\gamma_s (158 {\rm AGeV})$ is compatible
with the strangeness saturation factor reported in the previous
subsection emerging from a data set suitable for a centrality analysis. 
The open circle follows from an analysis of mid-rapidity
data of $\pi^\pm$, $K^\pm$, $\Lambda$ and $\bar \Lambda$. 
The change relative to results reported in the previous subsection
highlights the sensitivity on different data sets, 
if the number of analyzed hadron species is small.

\subsection{RHIC Energies}

Our analysis is restricted to the published data of $\pi^\pm$, $K^\pm$ and $p^\pm$
at $y \approx 0$ in centrality-binned collisions of Au + Au at $\sqrt{s_{NN}} = 130$ GeV 
\cite{PHENIX}. The upcoming data for further hadron species, including excited states, and 
data at $\sqrt{s_{NN}} = 200$ GeV will complete our picture on the systematics of hadron
multiplicities. 

\subsubsection{Centrality Dependence}

As for CERN-SPS energies, $\gamma_s$ increases with centrality, see right panel in Fig.~1.
The overall trend follows again the number of multiple participant collisions obtained
in a Glauber model calculation \cite{Jean_submit}. The most central collisions point to
strangeness saturation, in particular when feed-down from weak decays is included.

\subsubsection{Energy Dependence}

Fig.~2 suggests that increasing energy causes increased strangeness saturation.
However, this conclusion should be taken with caution, as Fig.~2 displays in one plot
analyses of $4\pi$ and $y \approx 0$ data and data sets comprising different hadron species. 
Assuming that at CERN-SPS the $4\pi$ data are sensible, while at RHIC only
$y \approx 0$ data should be analyzed within thermal models, then from Fig.~2
one does indeed derive the mentioned expectation: with increasing energy, $\gamma_s$
evolves towards saturation.

One observes in Fig.~2 that the thermal model delivers for $pp$ and $\bar pp$
collisions \cite{Becattini_Heinz} values of $\gamma_s$ compatible with peripheral
collisions. 

\begin{figure}[htb]
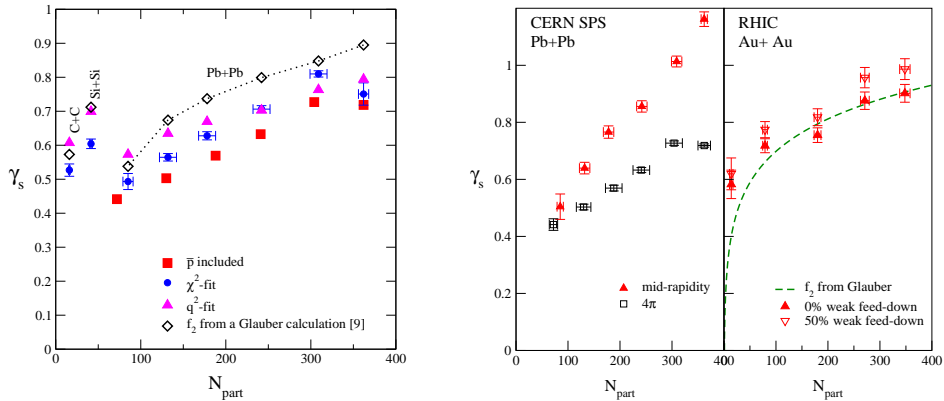

\vspace*{-.2cm} 
~\center
\epsfig{file=gammas_sys_size_mod.eps,width=5.3cm,angle=0} 
\hspace*{6mm}
\epsfig{file=gammas_mod.eps,width=6.3cm,angle=0} 
~\vskip -3mm
\caption[]{Left panel: System-size and centrality dependencies of 
$\gamma_s$, as extracted from centrality-binned Pb+Pb
\cite{Sikler,Blume} and central C+C 
and Si+Si data \cite{C-Si} under various fit conditions, 
assuming 50\% feeding from weak decays. 
Also shown is the fraction of participants which 
underwent multiple collisions, $f_2$.
Middle panel: Comparison of 
$\gamma_s$ extracted 
from mid-rapidity NA49 data \protect\cite{Sikler} 
with the results of our earlier analysis of NA49 $4\pi$-yields 
\protect\cite{Jean}. 
Right panel: $\gamma_s$ 
observed in Au+Au collisions as extracted from PHENIX data
\protect\cite{PHENIX};
$f_2$ from our Monte Carlo based Glauber model calculation \protect\cite{Jean_submit}. \\[-6mm]
}
\label{fig1}
\end{figure}

\begin{figure}[htb]
\begin{minipage}[b]{12.5cm}
\begin{minipage}[t]{5.5cm}
\vspace*{-.2cm}
~\center
\epsfig{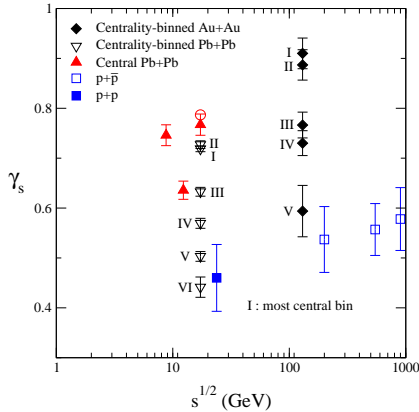} %
\vspace*{-.2cm}
\end{minipage}
\hfill
\begin{minipage}[t]{7.5cm}
\caption[]{Energy dependence of 
$\gamma_s$ extracted from central Pb+Pb collisions at 40, 80 and 
158 AGeV \protect\cite{NA49_coll1,NA49_coll2,NA49_ksi} 
together with the results of our earlier analysis of centrality-binned 
Pb+Pb collisions at 158 AGeV 
and Au+Au collisions at 
RHIC \protect\cite{Jean}. 
The open circle is extracted from mid-rapidity yields 
\protect\cite{NA49_coll1,NA49_ksi,Mischke}.
Results for $pp$ and $p \bar p$ collisions from \protect\cite{Becattini_Heinz}.}
\end{minipage}
\end{minipage}
\label{fig2}
\end{figure}

\section{Agreement with Data}

Up to now we have reported the system-size, centrality and energy dependence of $\gamma_s$.
The values of $T$ coincide approximately with the chiral (deconfinement)
transition temperature of $T \sim 165$ MeV, smoothly increasing from 150 MeV at beam energy
of 40 AGeV. $\mu_B$ (the baryon potential dominating $\vec \mu$) decreases from
350 to 220 MeV at CERN-SPS and drops to about 30 MeV at RHIC. The actual values depend
on the data samples analyzed. The agreement of data with multiplicities from the thermal
model is extremely good with the following exceptions:\\
$\bullet$ $\phi / K^+$ is underpredicted for peripheral collisions at beam energy of 158 AGeV,
even if the $\phi$ yields are included in the fit;\\
$\bullet$ $\Lambda + \Xi^0$ yields follow the data trend but
are in absolute normalization below data when using the
thermal parameters extracted from $\pi^\pm$, $K^\pm$ and $p^\pm$ yields at $\sqrt{s_{NN}} = 130$ GeV;
the $\Xi^-$ yields are at the lower end of the error bars.

To understand which trends in the data drive the change of $\gamma_s$ we exhibit in
Fig.~3 scaled multiplicities as a function of $N_{\rm part}$. A scaling
of $\bar p$ by $N_{\rm part}^{1.15}$ results in a flat distribution at beam energy of 
158 AGeV. The scaled $\pi^\pm$ is slightly decreasing, while $K^\pm$ increases.
It is the relative decrease of the "gap" between pion and kaon yields which drives
$\gamma_s$ towards unity.  
Similarly, at $\sqrt{s_{NN}} = 130$ GeV, the $p^\pm$ distributions are flat, as also
$K^\pm$, but the $\pi^\pm$ decrease with increasing $N_{\rm part}$. Again, the decreasing
"gap" between $\pi^\pm$ and $K^\pm$ drives the increase of $\gamma_s$. Note that the effects are small
and hardly visible in a log plot with unscaled yields.  

As an aside, we mention that we adjusted the parameters of the color glass model
\cite{Juergen_Larry} and obtain a good description of $\pi^\pm$ and $p^\pm$ yields,
while the model deviates in details from the $K^\pm$ data, see right
panel of Fig.~3.

\begin{figure}[htb]
\vspace*{-.01cm}
\epsfig{file=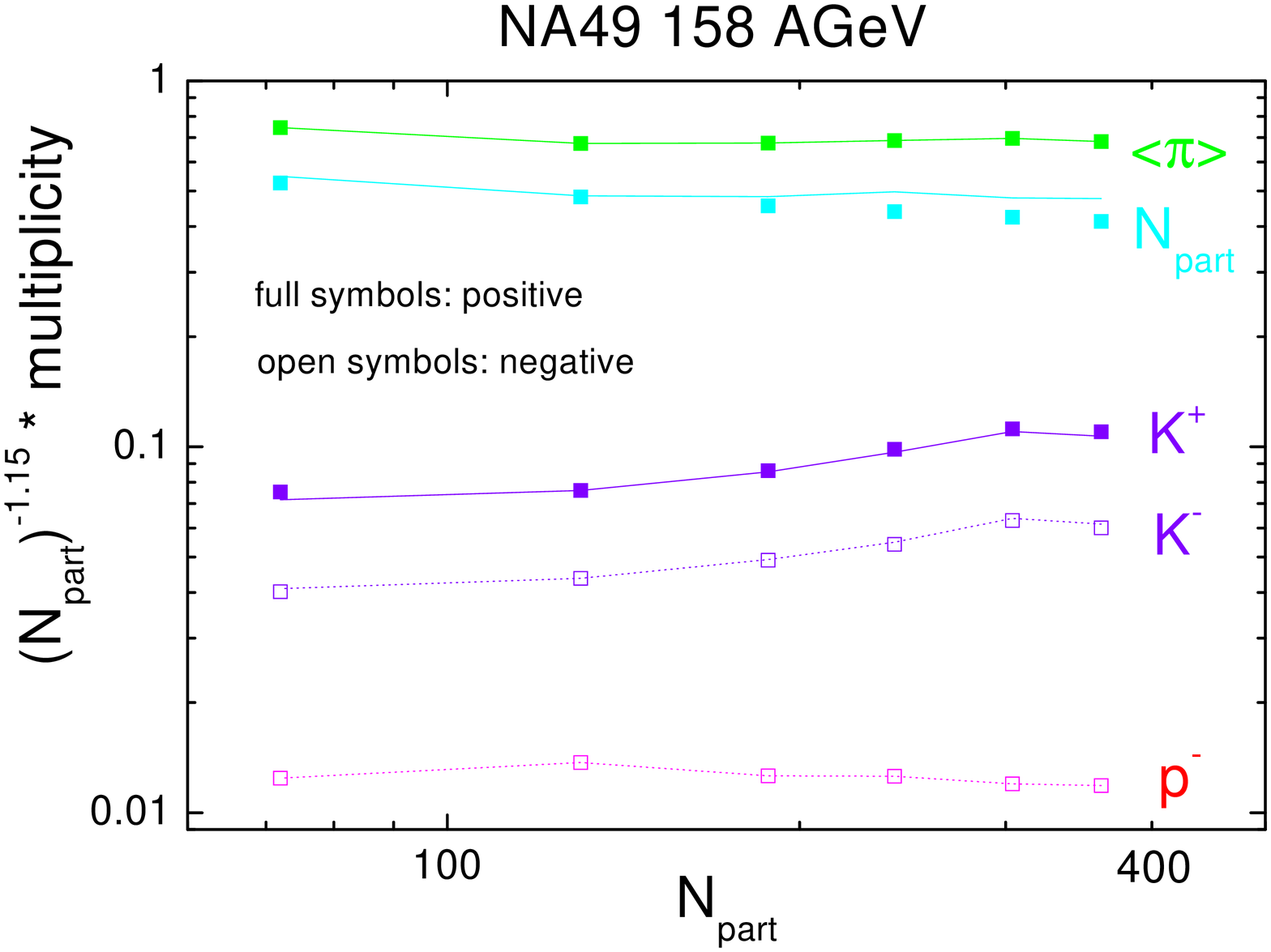,width=6.9cm,angle=0} 
\hspace*{-5mm}
\epsfig{file=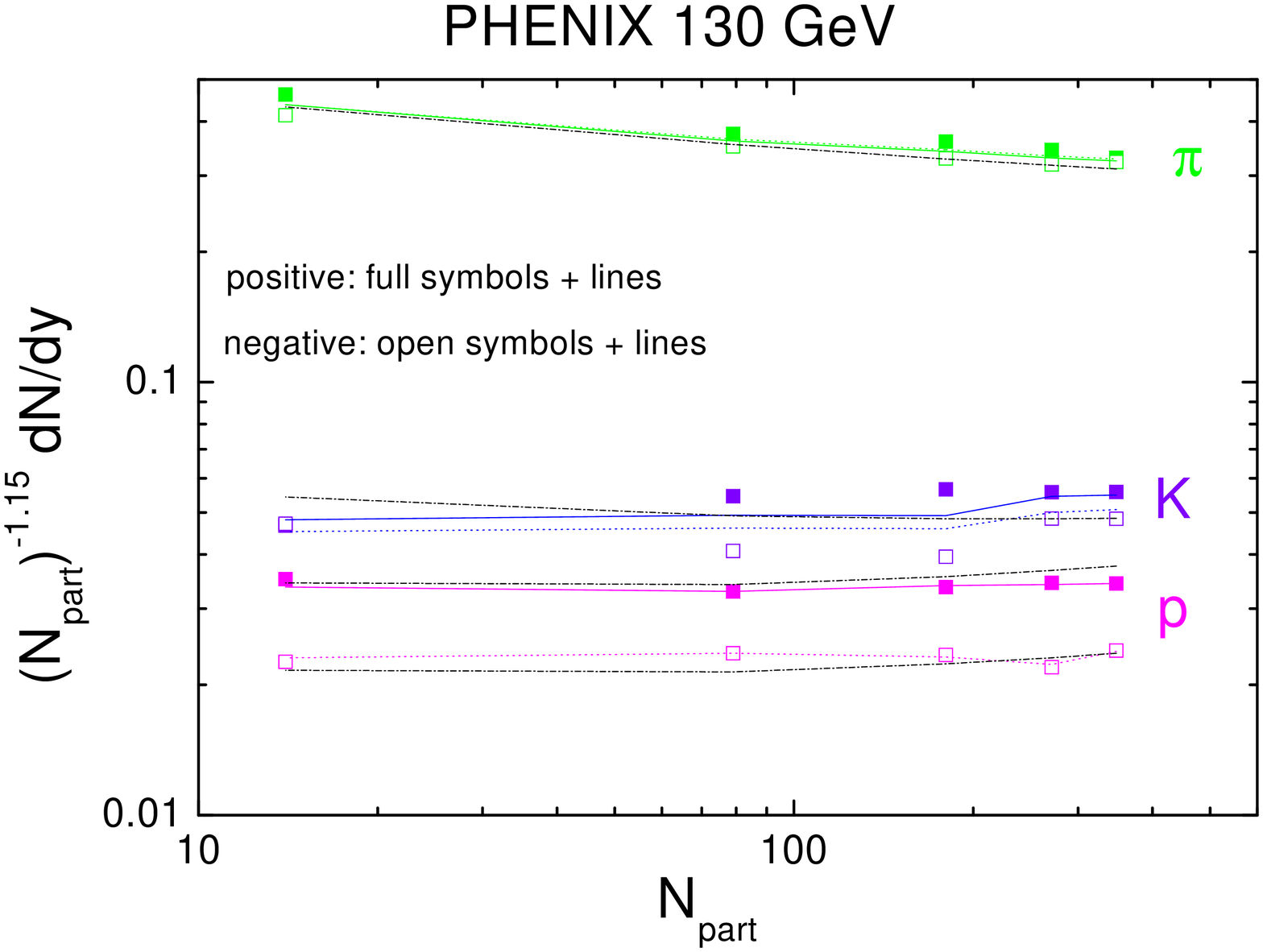,width=6.9cm,angle=0}
\vspace*{-.6cm}
\caption[]{Scaled hadron yields as a function of $N_{\rm part}$.
Left (right) panel: Pb + Pb (Au + Au), symbols: data \protect\cite{Sikler,Blume,PHENIX},
dotted and solid lines: results of the thermal model, 
dot-dashed curves: our results for the color glass model \protect\cite{Juergen_Larry}.}
\label{fig3}
\end{figure}

\section{Summary}

In summary we report here the dependence of the strangeness saturation factor, $\gamma_s$, 
on system size, centrality and energy. 
Most striking is the increase of $\gamma_s$ with centrality and the number of multiple
participant collisions. Details of the energy dependence of $\gamma_s$ are puzzling
and hampered by restricted data sets. Whenever possible,
one should compare the findings with results obtained from the richer data samples for
central collisions. Clearly, the present investigations should be
extended to include multi-strange baryons and resonances in the analysis of more complete
data sets. This will further test the applicability and predictive power of thermal models
of statistical hadronization in describing hadron multiplicities.


\section*{Note(s)} 
\begin{notes}
\item[a]
The work is supported by BMBF 06DR921 and 06DR121.
\item[b]
Visiting Fulbright Professor on leave of absence from 
the Brookhaven National Laboratory, Upton, NY, USA
\end{notes}

\vfill\eject
\end{document}